\documentclass[pre,aps,twocolumn,showpacs]{revtex4}
\usepackage{graphicx,epstopdf}
\usepackage{pdfpages}
\usepackage{dcolumn}
\usepackage{bm}
\usepackage{color}
\usepackage{amssymb,amsmath}
\begin{document}
\title{Tug-of-War in a Double-Nanopore System}

\author{Aniket Bhattacharya}

\altaffiliation[]
{Author to whom the correspondence should be addressed}
{}
\email{AniketBhattacharya@ucf.edu}

\author{Swarnadeep Seth}

\affiliation{Department of Physics, University of Central Florida, Orlando, Florida 32816-2385, USA}
\date{\today}
\begin{abstract}
We simulate a tug-of-war (TOW) scenario for a model double-stranded DNA
threading through a double nanopore (DNP) system. The 
DNA, simultaneously captured at both pores is subject to two equal and opposite forces
$-\vec{f}_L= \vec{f}_R$ (TOW), where $\vec{f}_L$ and $\vec{f}_R$ are the
forces applied to the left and the right pore respectively. Even though the net force on the DNA polymer $\Delta \vec{f}_{LR}=\vec{f}_L+ \vec{f}_R=0$, the mean first passage time (MFPT)
$\langle \tau \rangle$ depends on the magnitude of the TOW forces $ \left | f_L \right | = \left |f_R \right | = f_{LR}$. 
We qualitatively explain this dependence of $\langle \tau \rangle$ on $f_{LR}$ from the known
results for the single-pore translocation of a triblock copolymer
A-B-A with $\ell_{pB} > \ell_{pA}$, where $\ell_{pA}$ and  $\ell_{pB}$
are the persistence length of the A and B segments respectively. We 
demonstrate that the time of flight (TOF) of a monomer with index $m$
($\langle \tau_{LR}(m) \rangle$) from one pore to
the other exhibits quasi-periodic structure
commensurate with the distance between the pores $d_{LR}$. Finally, 
we study the situation where we offset the TOW biases so that 
$\Delta \vec{f}_{LR}=\vec{f}_L+ \vec{f}_R   \ne 0$, and
qualitatively reproduce the experimental result of the dependence of the MFPT
on $\Delta\vec{f}_{LR}$. We demonstrate that for a moderate bias,
the MFPT 
for the DNP system for a chain length $N$ follows the same scaling ansatz as that of for the
single nanopore, 
$\langle \tau \rangle  = \left( AN^{1+\nu}  + \eta_{pore}N \right)
\left(\Delta f_{LR}\right)^{-1}$, where $\eta_{pore}$ is the pore
friction, which enables us to estimate $\langle \tau \rangle $ for a long chain. 
Our Brownian dynamics  simulation studies provide fundamental insights and valuable information about the
 details of the translocation speed obtained from $\langle \tau_{LR}(m) \rangle$, and accuracy of the translation of the data obtained in the
time-domain to units of genomic distances.
\end{abstract}
\maketitle
\section{Introduction}
Nanopore (NP) sensing is a powerful approach for accurate, fast and
cost-effective detections of biomolecules, such as single and double
stranded DNA, peptides and proteins~\cite{Bashir}. 
For almost two decades, research in this area has spurred continued
and increased activities among broad
disciplines of sciences and engineering due to their direct impact on human
health and diseases. Unlike traditional
methods, which require molecular amplification, in a single NP (SNP) based method~\cite{Review}, a particular DNA segment is analyzed as the nucleotides make 
single file translocation through the NP. Since its original demonstration in
$\alpha$-hemolysin protein pore~\cite{Kasianowicz,Meller00,Meller01,Meller02}, NP translocation has been
studied in other biological NPs, nanopores in silicon nitride
membranes, and two dimensional (2D) materials, such 
multi-layered graphene NPs. Recently translocation of a DNA segment has
been extended in  double-NP systems after
being co-captured by both the pores~\cite{Dekker,TwoPore1,TwoPore2,Flossing,Langecker}. Compared to a single NP,
double, or multiple NPs detection methods with adjustable bias and feedback applied at each pore offer better control of the DNA. Different variations of this
concept, such as DNP separated by a nano-bridge~\cite{Cadinu1}, double-barrel NP~\cite{Cadinu2}, nanoscale pre-confinement~\cite{Briggs}, and
entropy driven TOW~\cite{Yeh} have
also been reported. \par
While translocation through a single
NP has been studied quite extensively theoretically, experimentally,
and using a variety of numerical and simulation strategies, theoretical
studies and modeling translocation in double or multiple NP system is
only directed to explain a specific experimental
system~\cite{Dekker}.\par
\vskip -0.0625 truecm
In this paper,  we report  Brownian dynamics (BD) simulation
studies of a coarse-grained (CG) model homopolymer translocating through a DNP
system. Our model DNP system {\em in silico}  is a simplified version of 
the experimental designs of DNP systems  
reported recently~\cite{TwoPore1,TwoPore2} in which we study 
the effects of chain stiffness $\kappa$, distance between the pores $d_{LR}$, the
magnitude of the biases  $\vec{f}_L$ and $\vec{f}_R$ on the MFPT
$\langle \tau \rangle $, and on the TOF of the
individual monomers to provide further details of various aspects
of translocation through a DNP. The CG model as described in the next
section does not require detail structures of the DNA and is
sufficient to answer the questions addressed in this paper. 
We have chosen $d_{LR} << L$, the contour length of the chain.
\begin{figure}[ht!]
\includegraphics[width=0.45\textwidth]{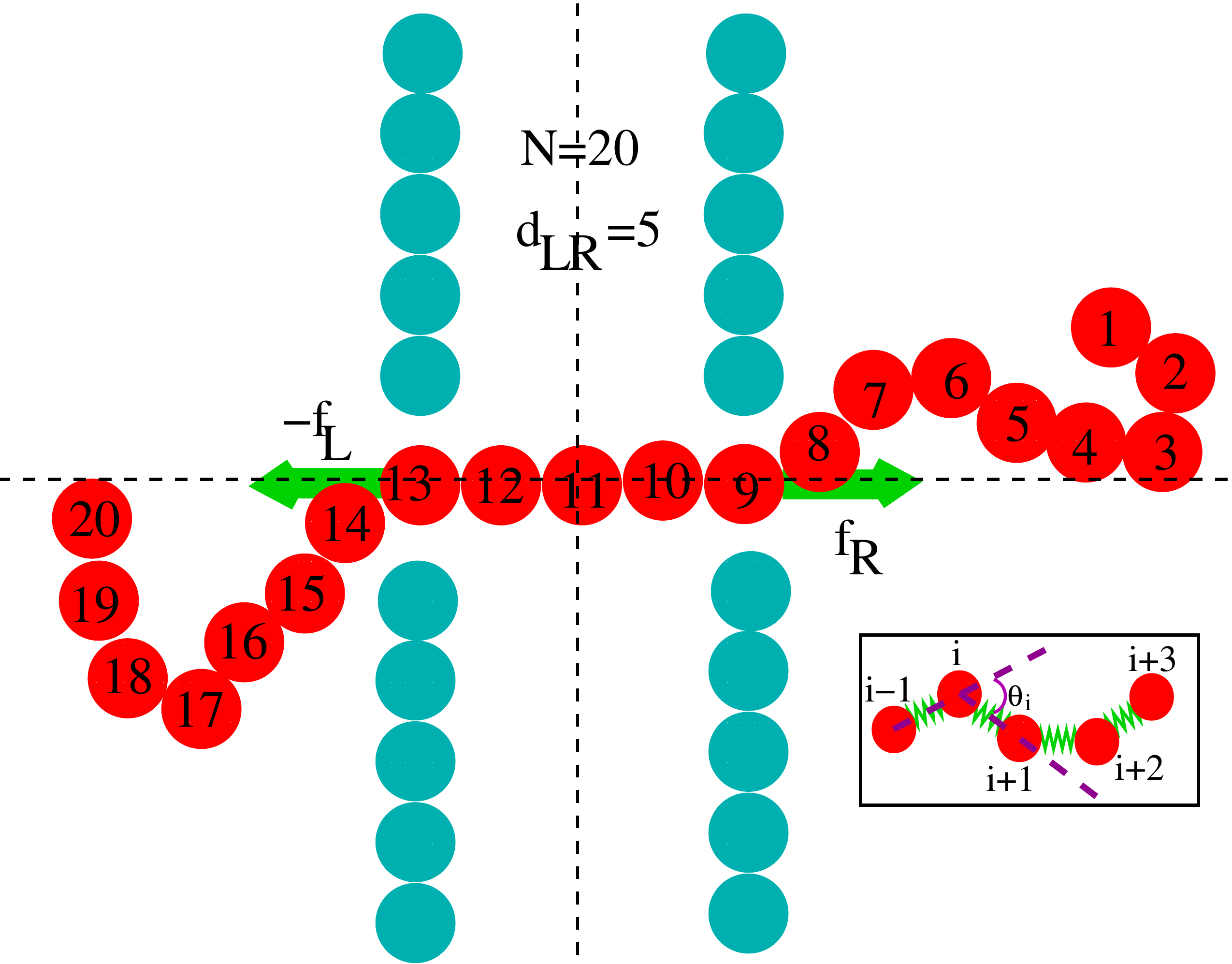}
 \caption{Schematic of BD simulation model for a chain of
  length $N=20$ (red).  Initial configuration of the chain in between the
  pores $d_{lR}$ apart is approximately
  a straight line. The immobilized walls on which the two pores are
  located are created by particles (teal) of the same
  diameter. External biases $\vec{f}_L$ and $\vec{f}_R$  are
  applied in each pore in opposite directions as shown. The box shows 
the details of the bead-spring model. The local persistence length
  $\ell_p(i) = -1/\ln \langle \cos \theta_i \rangle $ is calculated
  from the angle $\theta_i$ (Eqn.~\ref{local_lp})}. 
\label{Model}
\end{figure}
and often used the known results
from scaling theory of polymer translocation~\cite{Ikonen_EPL2013,Ikonen_JCP2012,Ikonen_PRE2012}, nonequilibrium tension propagation (TP)
theory of polymer translocation~\cite{Sakaue_PRE_2007}, and prior results for SNP translocation for a stiff chain
~\cite{Adhikari_JCP_2013} to explain the results for DNP translocation in this limit. These studies provide information 
to design new experiments with different parameter sets, develop a theoretical framework
that can be tested by additional simulation studies.
\section{Model}
\label{Model0}
 Our BD scheme is implemented on a bead-spring model of a polymer with
 the monomers interacting via an excluded volume (EV), a Finite Extension Nonlinear Elastic (FENE) spring potential, and a
 bond-bending potential enabling variation of the chain persistence length $\ell_p$ (Fig.~\ref{Model}).  
The model,  originally introduced for a fully flexible chain by  Grest and Kremer~\cite{Grest}, has been studied quite 
extensively by many groups using both Monte Carlo (MC) and various molecular dynamics (MD) methods~\cite{Binder_Review}. 
Recently we have generalized the model for a semi-flexible chain and studied both equilibrium and dynamic properties~\cite{Adhikari_JCP_2013,Huang_EPL_2014a}. 
Comparison of our BD results with those obtained for very 
large self-avoiding chains on a square lattice reveals robustness of the model for certain universal aspects, {e.g.}, 
scaling of end-to-end distance and transverse fluctuations~\cite{Huang_EPL_2014a, Huang_EPL_2014b,Huang_JCP_2014,Huang_JCP_2015}. 
Using our BD scheme for confined stiff polymers in nanochannels we have demonstrated and verified the existence of Odijk deflection length 
$\lambda \sim (\ell_pD^2)^{1/3}$~\cite{Huang_JCP_2015}. More recently we compared the evolution of the density profile along the nanochannel axis obtained from 
the BD simulation with those obtained from an approach using Nonlinear Partial differential equation~\cite{Polymers2016} with excellent agreement showing the 
applicability of the BD simulation method  to study nonequilibrium dynamics of confined polymers. The BD simulation provides detailed
picture of how a stiff chain folds into a series of nested loops when
pushed by a nanodozer~\cite{MM2018}. Last but not the least we have
used the same model earlier to address various problems in SNP
translocation with
success~\cite{Kaifu_PRL,Bhattacharya_EPJE,Bhattacharya2010,Bhattacharya_Proceedia}. The
successes of these prior studies explaining a variety of phenomena provide
assurance that the BD simulation studies will provide useful informations and insights toward a fundamental
understanding of polymer translocation through a model DNP system. 
\par
The EV interaction between any two monomers is given by a short range Lennard-Jones (LJ) potential
\begin{eqnarray}
U_{\mathrm{LJ}}(r)&=&4\epsilon \left[{\left(\frac{\sigma}{r}\right)}^{12}-{\left(\frac{\sigma}
{r}\right)}^6\right]+\epsilon, \;\mathrm{for~~} r\le 2^{1/6}\sigma; \nonumber\\
        &=& 0, \;\mathrm{for~~} r >  2^{1/6}\sigma.
\label{LJ}
\end{eqnarray}
Here, $\sigma$ is the effective diameter of a monomer, and 
$\epsilon$ is the strength of the LJ potential. The connectivity between 
neighboring monomers is modeled as a FENE spring with 
\begin{equation}
U_{\mathrm{FENE}}(r)=-\frac{1}{2}k_FR_0^2\ln\left(1-r_{ij}^2/R_0^2\right).
\label{FENE}
\end{equation}
Here $r_{ij}=\left | \vec{r}_i - \vec{r}_j \right|$ is the distance
between the consecutive monomer beads $i$ and $j=i\pm1$ at $\vec{r}_i$
and $\vec{r}_j$, $k_F$ is the spring constant and $R_0$
is the maximum allowed separation between connected monomers. 
The chain stiffness $\kappa$ is introduced by adding an angle dependent three body interaction term between successive bonds 
as (Fig.~\ref{Model}) 
\begin{equation}
U_{\mathrm{bend}}(\theta_i) = \kappa\left(1-\cos \theta_i\right) 
\end{equation}
Here $\theta_i$ is the angle between the bond vectors 
$\vec{b}_{i-1} = \vec{r}_{i}-\vec{r}_{i-1}$ and 
$\vec{b}_{i} = \vec{r}_{i+1}-\vec{r}_{i}$, respectively, as shown in Fig.~\ref{Model}. The strength 
of the interaction is characterized by the bending rigidity $\kappa$
associated with the $i^{th}$ angle $\theta_i$.
For a homopolymer chain the bulk persistence length $\ell_p$ of the chain in
two dimensions (2D) is
given by~\cite{Landau}
\begin{equation}
  \ell_p/\sigma = 2\kappa/k_BT.
  \label{lp_bulk}
  \end{equation}
\par 
Each of the two purely repulsive walls consists of one mono-layer
(line) of immobile LJ particles of the same diameter $\sigma$ of the
polymer beads symmetrically placed at $\pm \frac{1}{2}d_{LR}$. 
The two nanopores are created by removing two particles at the center
of each wall. 
We use the Langevin dynamics with the following equations of motion for the i$^{th}$ monomer 
\begin{equation}
m \ddot{\vec{r}}_i = -\nabla (U_\mathrm{LJ} + U_\mathrm{FENE} + U_\mathrm{bend} \\
                                            + U_\mathrm{wall}) -\Gamma \vec{v}_i + \vec{\eta}_i . 
                                          \label{langevin}                                          
\end{equation}

Here $\vec{\eta} _ i (t)$ is a Gaussian white noise with zero mean at temperature $T$, and 
satisfies the fluctuation-dissipation relation in $d$ physical
dimensions (here $d=2$):
\begin{equation}
< \, \vec{\eta} _ i (t) 
\cdot \vec{\eta} _ j (t') \, > = 2dk_BT \Gamma \, \delta _{ij} \, \delta (t 
- t ').
\end{equation}
We express length and energy in units of $\sigma$ and $\epsilon$, respectively. 
The stiffness parameter $\kappa$ is expressed in units of $\epsilon$, 
and the parameters for the FENE potential in Eq.~(\ref{FENE}), $k_F$ and 
$R_0$, are set to $k_F = 30 \epsilon/\sigma$ and $R_0 = 1.5\sigma$, respectively. 
The friction coefficient and the temperature are set to 
$\Gamma = 0.7\sqrt{m\epsilon/\sigma^2}$, $k_BT/\epsilon = 1.2$,
respectively. The force is measured in units of $k_BT/\sigma$.
The numerical integration of Equation~(\ref{langevin}) is implemented using the algorithm introduced by Gunsteren and Berendsen~\cite{Langevein}.   Our previous experiences with BD simulation suggests that for a time step $\Delta t = 0.01$ these parameters values produce stable trajectories over  a very long period of time and do not lead to unphysical crossing of a bond by a monomer~\cite{Huang_JCP_2014,Huang_JCP_2015}.  The average bond length stabilizes at $b_l = 0.971 \pm 0.001$ with negligible fluctuation regardless of the chain size and rigidity~\cite{Huang_JCP_2014}. We have used a Verlet neighbor list~\cite{Allen} instead of a link-cell list to expedite the computation.

\section{Simulation Results}
We carried out simulations for chain lengths $N  = L/\sigma = 64$, 96, 128,
192, 256, 320, and 384 (where $L$ is the corresponding chain contour length) for various chain stiffness $\kappa = 0, 16, 32, 64$,
as well as for several distances between the pores $d_{LR}$ for consistency 
checks, but show only limited set of results $d_{LR}/\sigma = 16$ and 
32 and mostly for $\kappa = 8$ and 16. 
The simulation results are averaged over at least 2000 initial conditions. 
For exact tug-of-war situation and for small biases the computations 
can be prohibitively large compared to even a weakly biased situation. \par
The starting point of our study is a homopolymer already co-captured by both the pores and placed 
symmetrically with the number of beads at the left side of the
left pore ($n_L$) and at the right side of the right pore ($n_R$) are
the same~(Fig.~\ref{Model}), so that $n_L+n_R+n_m = 2n_L+n_m = N$, where $n_m$ represents the number of monomers in between the 
two pores. Our initial
configuration is a straight chain with $d_{LR} = n_m\sigma$, which is then equilibrated with BD simulation time about 5 times the Rouse relaxation time $\tau_{eq}\propto N^{1+2\nu}$ 
keeping the two beads, located inside the left and the right pore clamped, where $\nu=0.75$
is the Flory exponent in two dimensions~\cite{Rubinstein}.
In simulation the local chain persistence length is
calculated from
  $\langle \cos \theta_m\rangle $ via
\begin{equation}
\ell_p(m) = -\ln\frac{1}{\left( \langle \cos \theta_m \rangle \right)},
\label{local_lp}
\end{equation}
where $m$ is the monomer index. 
Previously we have checked that the two definitions (Eqns. ~\ref{lp_bulk} and ~\ref{local_lp}) 
  become equivalent for a free homopolymer chain and
  that for a heteropolymer chain, the calculation of the $\langle \cos \theta _i
  \rangle $ provides the correct way to determine the local chain persistence
  length~\cite{Huang_EPL_2014a}. 
  After the polymer chain is equilibrated with beads inside the left and
right pores at the clamped positions, the chain is allowed to translocate with biases applied at the left and the 
right pores as shown in Fig.~\ref{Model}. We consider both the cases
$\vec{f}_L+\vec{f}_R = 0 $ and $\vec{f}_L+\vec{f}_R \ne 0 $.
\subsection{TOW - local chain persistence length $\ell_p(m)$}
\begin{figure}[ht!]
\includegraphics[width=0.23\textwidth]{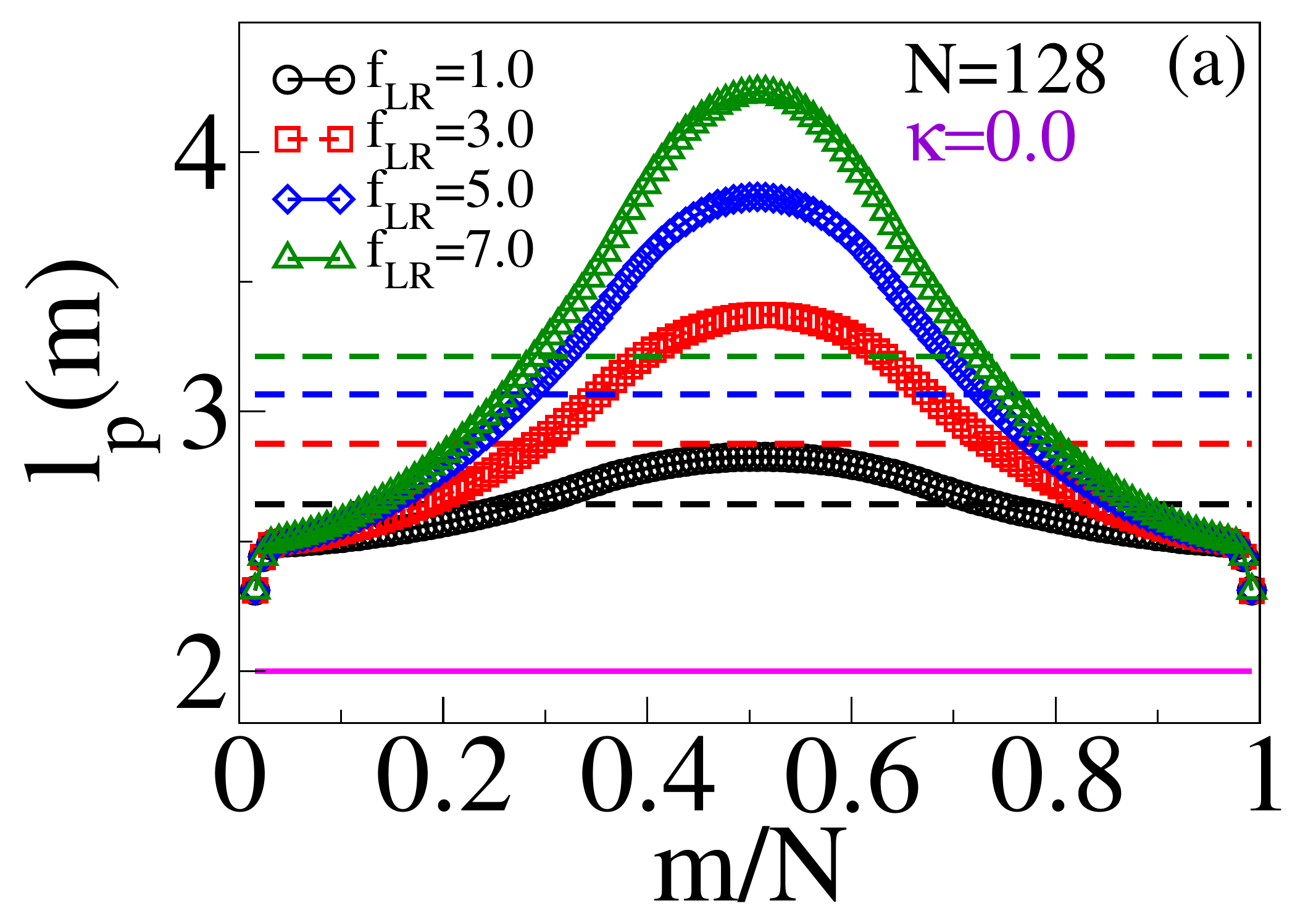}
\includegraphics[width=0.23\textwidth]{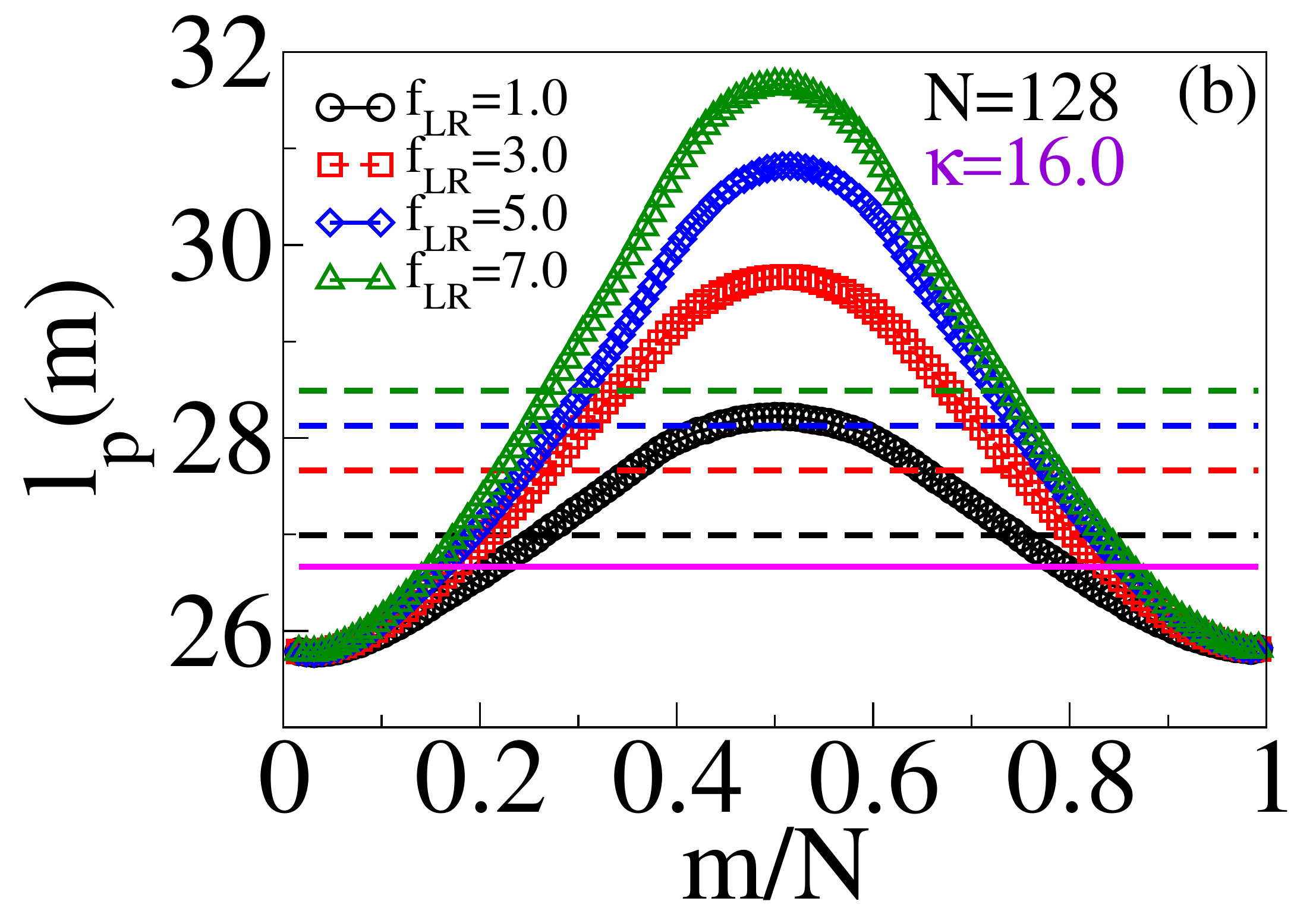}\\
\includegraphics[width=0.23\textwidth]{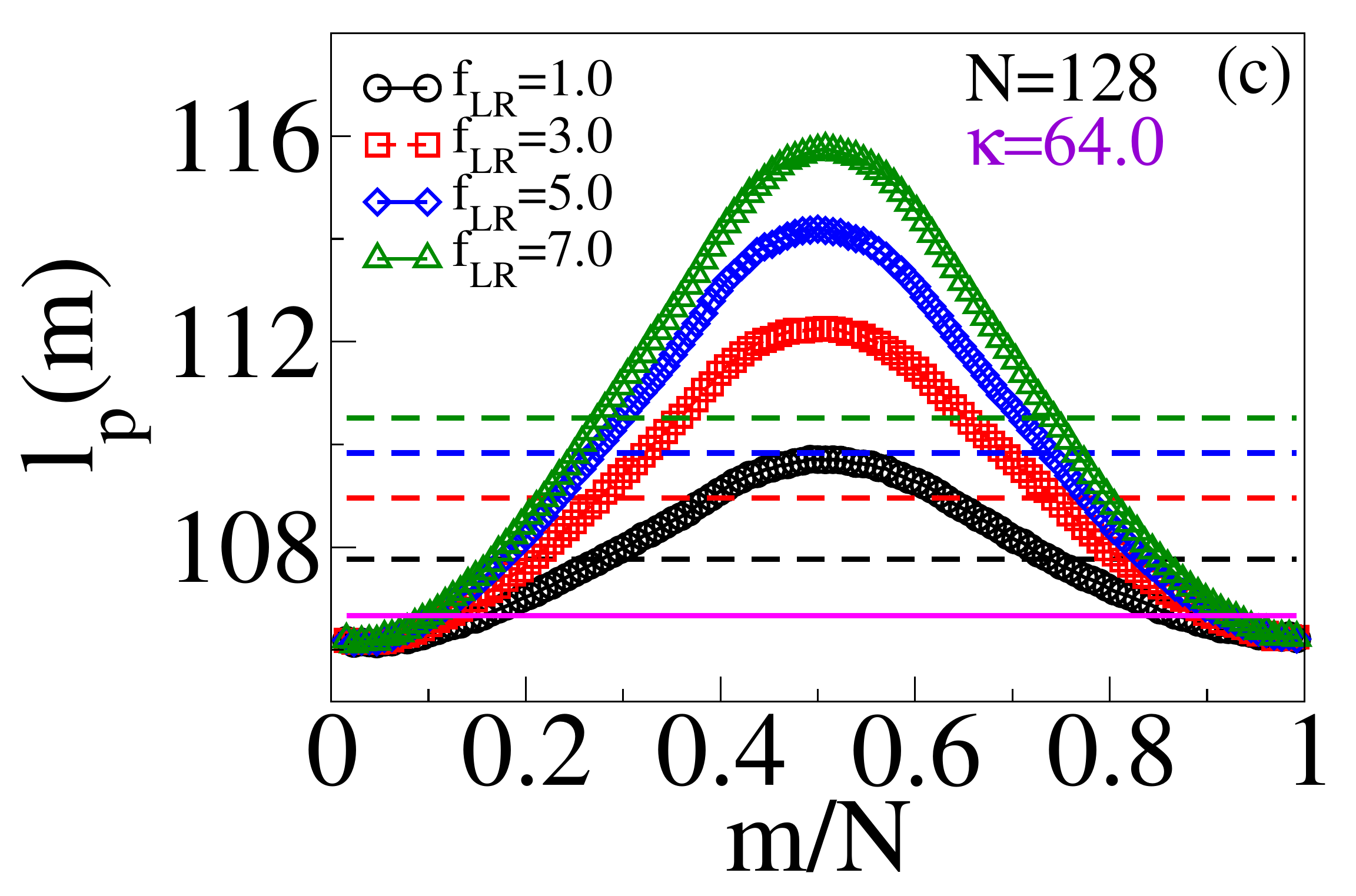}
\includegraphics[width=0.23\textwidth]{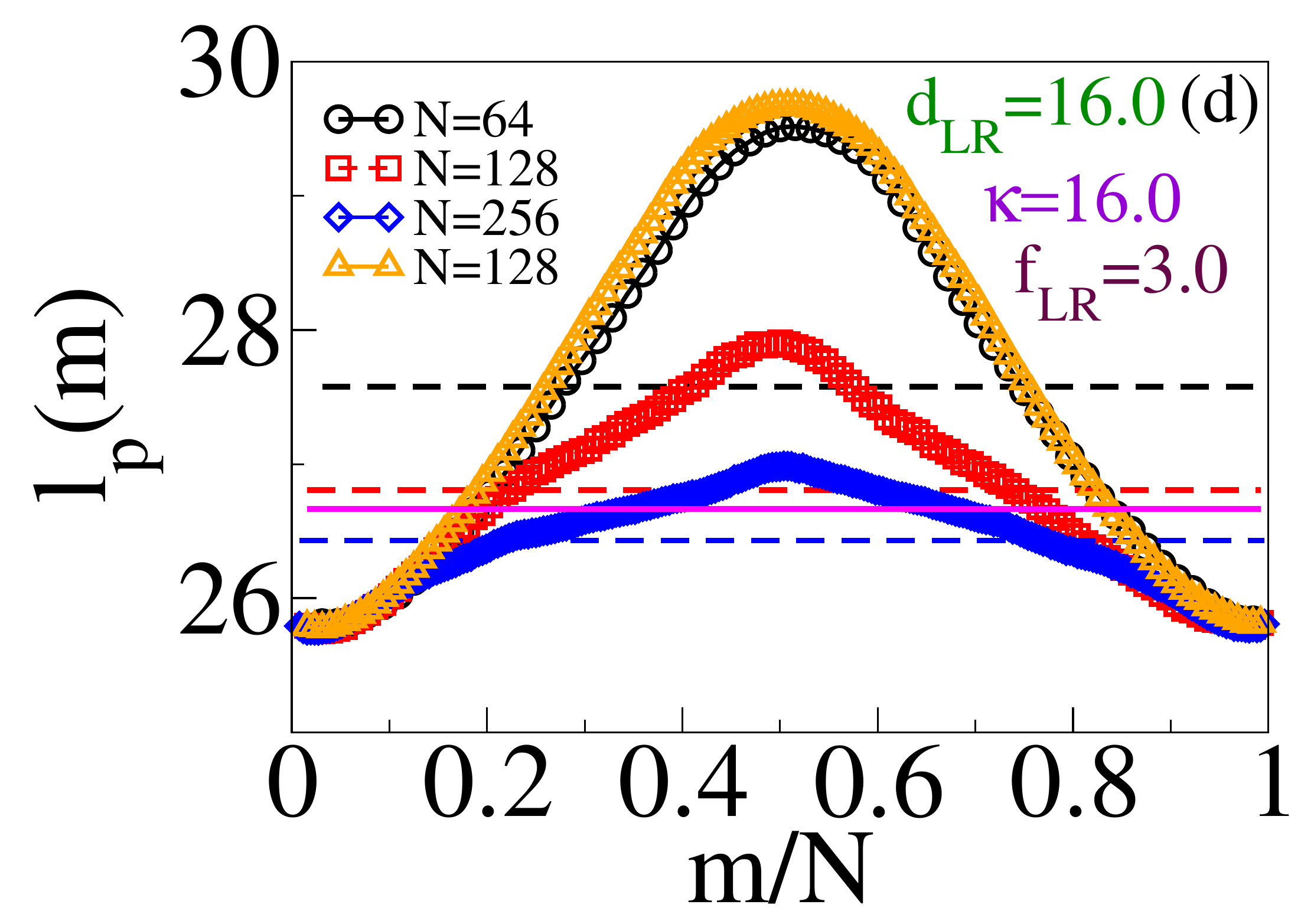}
\caption{\label{lp}\small The local chain persistence length $\ell_p(m)$ as a 
  function of reduced monomer index $m/N$. (a) for 
  $\kappa = 0.0$, (b) $\kappa = 16.0$, and (c) $\kappa =
  64.0$. The black circles, red squares, blue diamonds, and green 
  triangles correspond to TOW bias $f_{LR} =$1, 3, 5, and 7 
  respectively.  Fig.~(d) shows the effect of an increase in chain length $N$; 
  black circles, red squares, and blue diamonds are for $N=$64, 128, 
  and 256 respectively. The orange triangles and the black squares compare a case 
  where $d_{LR}/N = 16/64 = 32/128= \frac{1}{4}$ for two chain lengths. Each 
  dashed line is the average $\ell_p$ for the same color. In (a)-(c), the solid 
  purple line represents $\ell_{p0}=2\kappa/k_BT$ for the 
  unconstrained chain.}
\end{figure}

~First we study
the TOW situation where the chain is subject to two
equal and opposite forces $\vec{f}_L$
and $\vec{f}_R$ (Fig.~\ref{Model}) at the left and right
pores. 
Since the net force is zero, the chain executes diffusive motion until it
translocates (exits) either through the left or through the right
pore. However, unlike an unbiased translocation in a SNP, the persistence length of the chain segment in between
the two pores becomes larger due to the presence of TOW forces.
This causes the 
effective local persistence length to be a function of the monomer index
$m$, so that the average effective persistence length of the entire
chain becomes 
larger as shown in Fig.~\ref{lp}. We observe that the effect is most
prominent for $\kappa = 0$
(Fig.~\ref{lp}(a)), which will be relevant if similar experiments are
performed for a single stranded DNA. 
Figs.~\ref{lp}(a)-(c) for $\kappa = 0$, 16, and 64 look qualitatively similar, however, the scales are
very different. Evidently, the effect is less pronounced for a stiffer
chain as the relative increase in the chain persistence length is less for the same TOW forces $f_{LR}$. \par
\par
\subsection{Tug-of-war and MFPT}
One then wonders how does this variation in chain  
persistence length affect the MPFT ? 
 Fig.~\ref{tow-lp}(a) shows the variation of MFPT as a function of the
TOW forces $f_{LR}$. Consistence with Fig.~\ref{lp} we observe a
noticeable increase in $\langle \tau \rangle$ for $\kappa = 0$, and a relatively small 
increment  for chains with $\kappa = 16$ and 64. 
The result can be understood from a prior result for the single pore
translocation~\cite{Adhikari_JCP_2013},
where it has been shown that the $\langle \tau \rangle $ increases
with increasing chain stiffness $\kappa$. For the DNP, the 
TOW forces make the chain segment between the pores stiffer. The
relative degree of increase in persistence length depends on the original
stiffness $\kappa$ of the chain, measured in terms of
the ratio $(\ell_p(N/2) -\ell_p(1))/\langle \ell_p \rangle $ is about
80\%, 20\%, and 7\% for $\kappa = 0$, 16 and 64 respectively. This
explains why for the same set of $f_{LR}$, the relative increase in chain
stiffness for the segment in between the pores is less significant for
$\kappa = 16$ and 64, 
compared to $\kappa = 0$.  It is worthwhile to note that for longer chains when $d_{LR}/L
\rightarrow 0$, the entropic forces of the free segments on either
side of the pores become the dominant forces. Thus, it is conceivable
that the tiny slope (Fig.~\ref{tow-lp}(a)) observed for $\kappa =
16$ and 64) of the TOW for chains of
size $N=128$ is a finite size effect. For longer chains $\langle
\tau \rangle$ will have no dependence on the $TOW$ forces $f_{LR}$.
\par
\subsection{TOW  and a model triblock copolymer A-B-A}
\begin{figure}[ht!]
\includegraphics[width=0.4\textwidth]{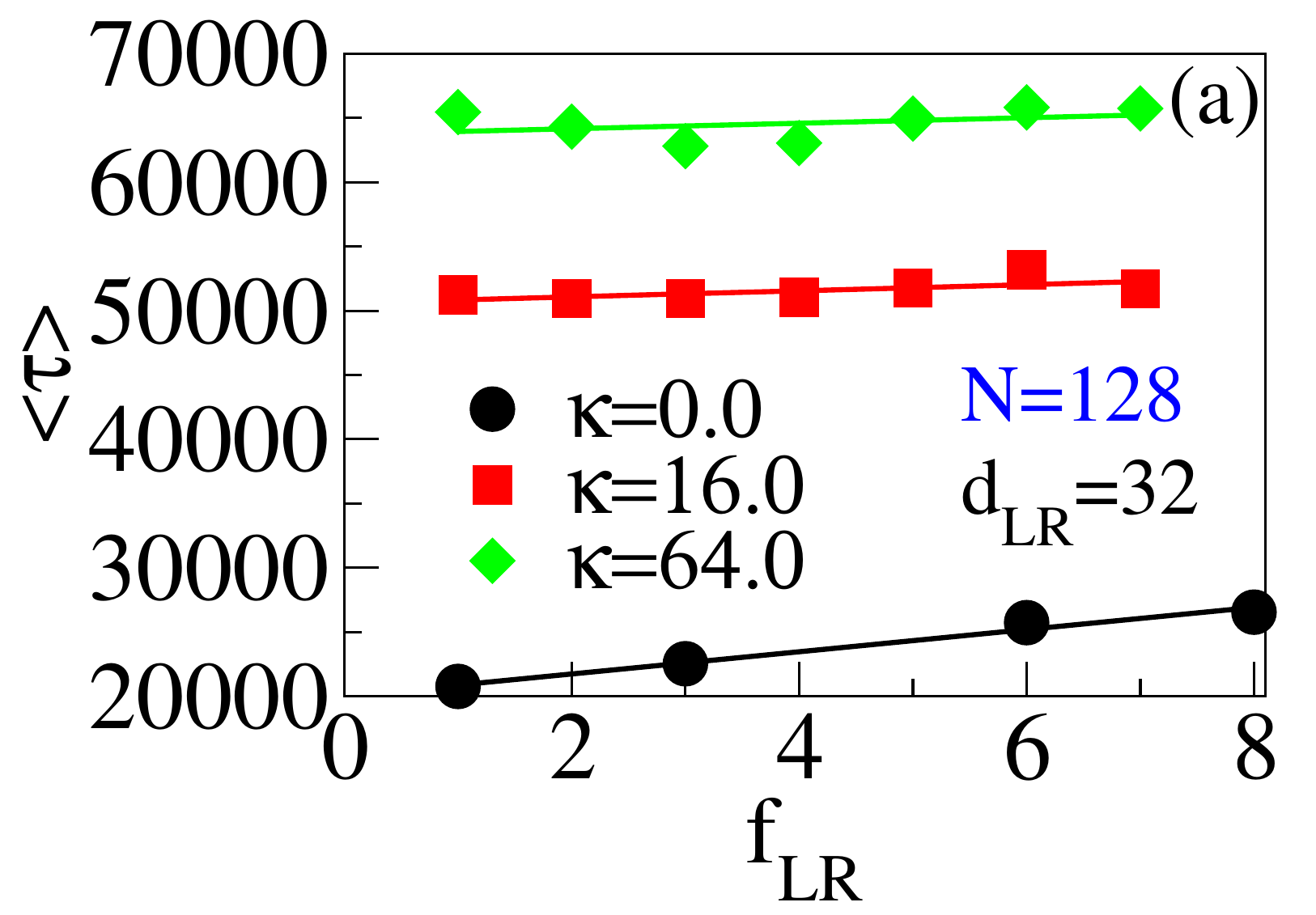}
\includegraphics[width=0.4\textwidth]{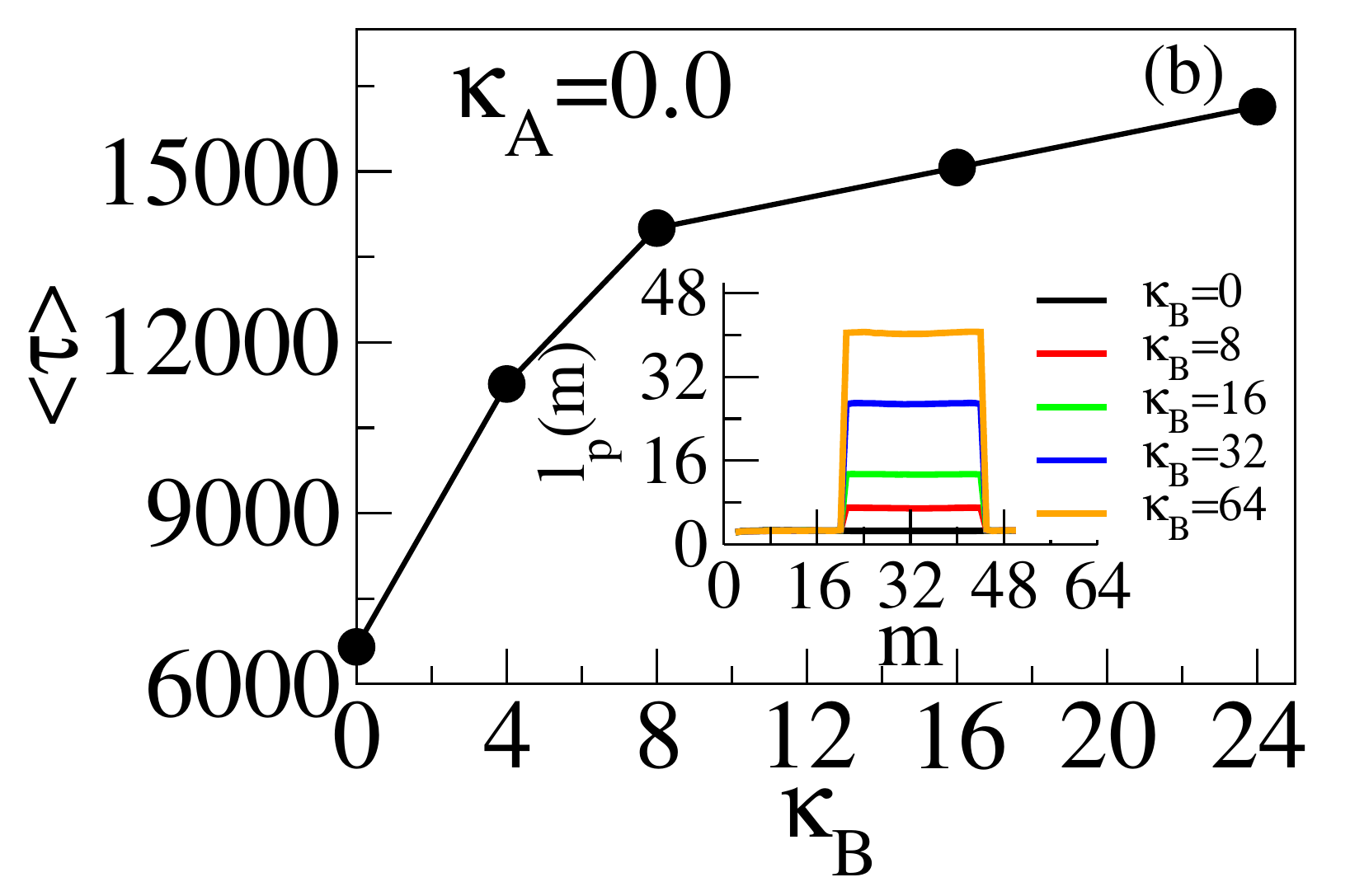}
\caption{\label{tow-lp}\small (a) MFPT as a function of TOW forces $f_{LR}$
for $\kappa = 0$ (black circles), $\kappa = 16$ (red squares), and 
$\kappa = 64$ (green diamonds). The straight lines through the points
are least square fits through the simulation data. 
(b) MFPT for a triblock copolymer chains A-B-A ($\kappa_A=0.0$)  as a function of $\kappa_B$ 
 The inset shows the corresponding $\ell_p(m)$ for $\kappa_B=
   0$, 8,16, 32, and 64 respectively.}
\end{figure}

We validate our interpretation by performing a separate set of
simulation. We set $f_L = f_R = 0$ and study the translocation of a
triblock copolymer of the form  ({\em flexible-stiff-flexible})
such that  $2n_A+n_B = N$, and $n_B \sigma = d_{LR}$,
where $n_A$ and $n_B$ are the length segments of the A and B segments
respectively. We choose the chain stiffness $\kappa_B > \kappa_A$.
 Keeping $\kappa_A$ constant we calculate the MFPT as an increasing
 function of $\kappa_B$. As expected, the translocation time increases 
as a function of the chain stiffness $\kappa_B$
(Fig.~\ref{tow-lp}(b)). 
We should mention that analogy is valid on an average.
The difference with the A-B-A copolymer  and 
the a homopolymer subject to a TOW 
is that in the former case, the persistence lengths along the chain are fixed, while for 
the TOW, it is depends on the location of the chain segment with
respect to the two pores. A translocating 
segment will have an increased persistence length while 
residing in the
region in between the pores. 
Thus on an average, in a TOW situation, the magnitude of  equal and 
opposite biases at each pore location 
affects the translocation time for a homopolymer chain.
(Fig.~\ref{tow-lp}(b)). 
The dependence of MFPT on 
$\Delta\vec{f}_{LR}$ 
will be relevant for experiments done with a single stranded DNA (ssDNA).  \par 

\subsection{TOW and the TOF $\langle \tau_{LR}(m) \rangle$}
\begin{figure}[ht!]
\includegraphics[width=0.45\textwidth]{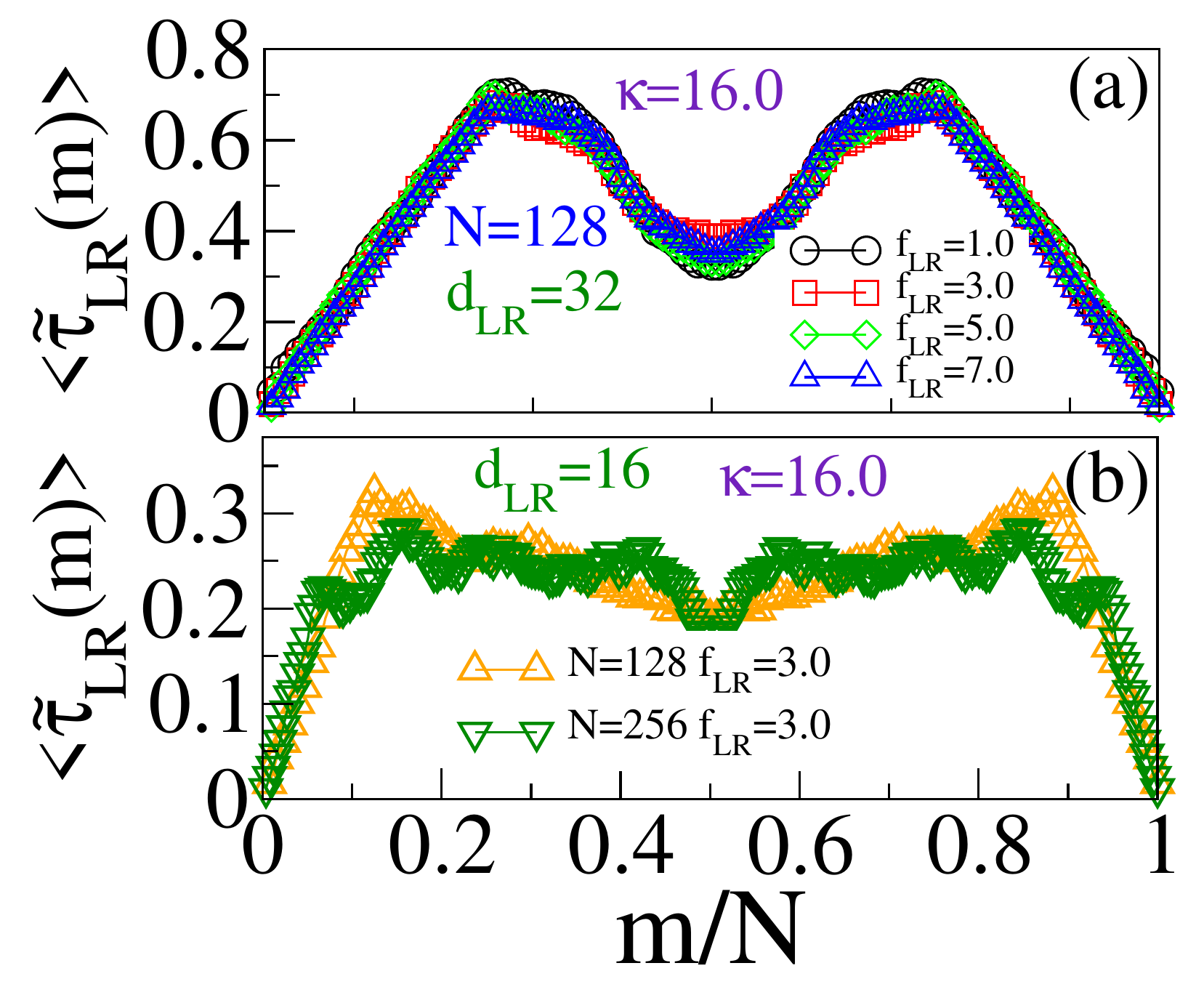}
\caption{\label{tof}\small The normalized TOF $\langle \tilde{\tau}_{LR}(m)\rangle$
as a function of the reduced monomer 
index $m/N$. (a) The 
  top curves are for $N=128$ for four different magnitudes of  
the TOW forces $f_{LR} = 1$ (black circles),  $f_{LR} = 3$ (red  
squares) and green diamonds ($f_{LR} = 3$), and $f_{LR} = 7$ (blue triangles) respectively.
(b). The bottom two curves are for $d_{LR}=16$ for chain 
  lengths $N=128$ (yellow triangles)  and $N=256$ (green triangles) respectively.}
\end{figure}

BD simulation provides detail information about the segmental translocation 
 process. One of the key questions in a TOW situation is  how long
 does a monomer with index $m$ take to cross the region in between
 the pores during the translocation (denoted as $\langle \tau_{LR}(m)
 \rangle$) ? Experimentally this quantity 
 is measured repeatedly in a DNA flossing experiment~\cite{Flossing}. The TOF $\langle \tau_{LR}(m) \rangle$ should be contrasted with the MFPT $\langle
 \tau \rangle$, which is the average total time of translocation for
 the entire chain. Since the TOF  $\langle \tau_{LR}(m) \rangle$ can be measured experimentally~\cite{TwoPore2,Flossing}, 
it can provide further informations. The dependence of the normalized
TOF defined as 
\begin{equation}
    \langle \tilde{\tau}_{LR}(m) \rangle =\frac{\langle \tau_{LR}(m)
      \rangle}{\langle \tau_{LR} \rangle}
    \end{equation}
\label{tof_eqn}
on $f_{LR}$ and $d_{LR}$ is
shown in Fig.~\ref{tof}, where
\begin{equation}
\langle \tau_{LR} \rangle  =\sum_m \langle \tau_{LR}(m) \rangle.
\end{equation}
Please note that the plot is made symmetric by combining
 the data from the left and right translocation (we checked that the
 data looks 
 statistically similar with 50\%  translocation from left to right and
 {\em vice-versa}). In experiments~\cite{TwoPore1,TwoPore2}, the ratio
 $d_{LR}/L << 1$. We also show some results for $d_{LR}/L = 0.25$
 to understand the limit $d_{LR}/L << 1$ better. \par

 Fig.~\ref{tof}(a)  show significant variations in $\langle \tilde{\tau}_{LR}(m) \rangle$. The almost linear decrease for $m/N \ge 0.75$ (or rise $m/N
 < 0.25$) corresponds to the last/first  32 (25\%) monomers
 exiting through the right/left pore when they are subject to only one of the
 TOW forces. The monomers $m \simeq N/2$ can have a nonzero $\langle \tilde{\tau}_{LR}(m) \rangle$ if
 they first travel either to the left/right pore and, then finally
 exit. Thus the $\langle \tilde{\tau}_{LR}(m) \rangle$  is minimum at $m=N/2$. 
 The monomers which follow the central
 monomer has to have an increased $\langle \tilde{\tau}_{LR}(m) \rangle$ until $m/N \simeq
 d_{LR}/N\sigma$. This explains the shape of the four $\langle \tilde{\tau}_{LR}(m) \rangle$
 for $d_{LR}/L = 32/128 = 1/4$ in Fig.~\ref{tof}(a).
One observes that the shape of the $\langle \tilde{\tau}_{LR}(m) \rangle$  is independent of the
magnitude of the TOW force $f_{LR}$ provided that the value of the
chain stiffness is high enough compared to other parameters of the
system.\\

 What happens in the limit  $d_{LR}/L \rightarrow 0$
 ? The experiments are done in this limit. This limit can be predicted
 from the shape of the two curves of Fig.~\ref{tof}(b). Here we
 plot the corresponding $\langle \tilde{\tau}_{LR}(m) \rangle$ for $d_{LR}/L = 16/128 = 0.125$
 and  $d_{LR}/L = 16/256 = 0.0625$ so that $d_{LR} << L$. We
 notice similar feature at the end and at the center. However, for $N=256$
 we observe a reasonably
 flat $\langle \tilde{\tau}_{LR}(m) \rangle$, albeit with  a small amplitude periodic oscillation
 in units of $d_{LR}/L$ for monomers satisfying $0.5d_{LR}/L \le m/N <
 \left(L-d_{LR}\right)/L$. This we believe is due to different environment a monomer
 encounters as it enters from {\em the region located at the left side of the left pore
 $\rightarrow$ the region in between the pores  $\rightarrow $ the
 region located at the right side of the right pore}
 of width $d_{LR} \simeq m\sigma$. The fine structure of the
 $\langle \tilde{\tau}_{LR}(m) \rangle$ can possibly be detected
from DNA flossing experiments reported
 recently~\cite{Flossing}, where the current blockade due to a known genomic length segment tagged by
 proteins is measured repeatedly by altering the bias between the pores.
 \subsection{Translocation with a net bias}
\begin{figure}[ht!]
\includegraphics[width=0.45\textwidth]{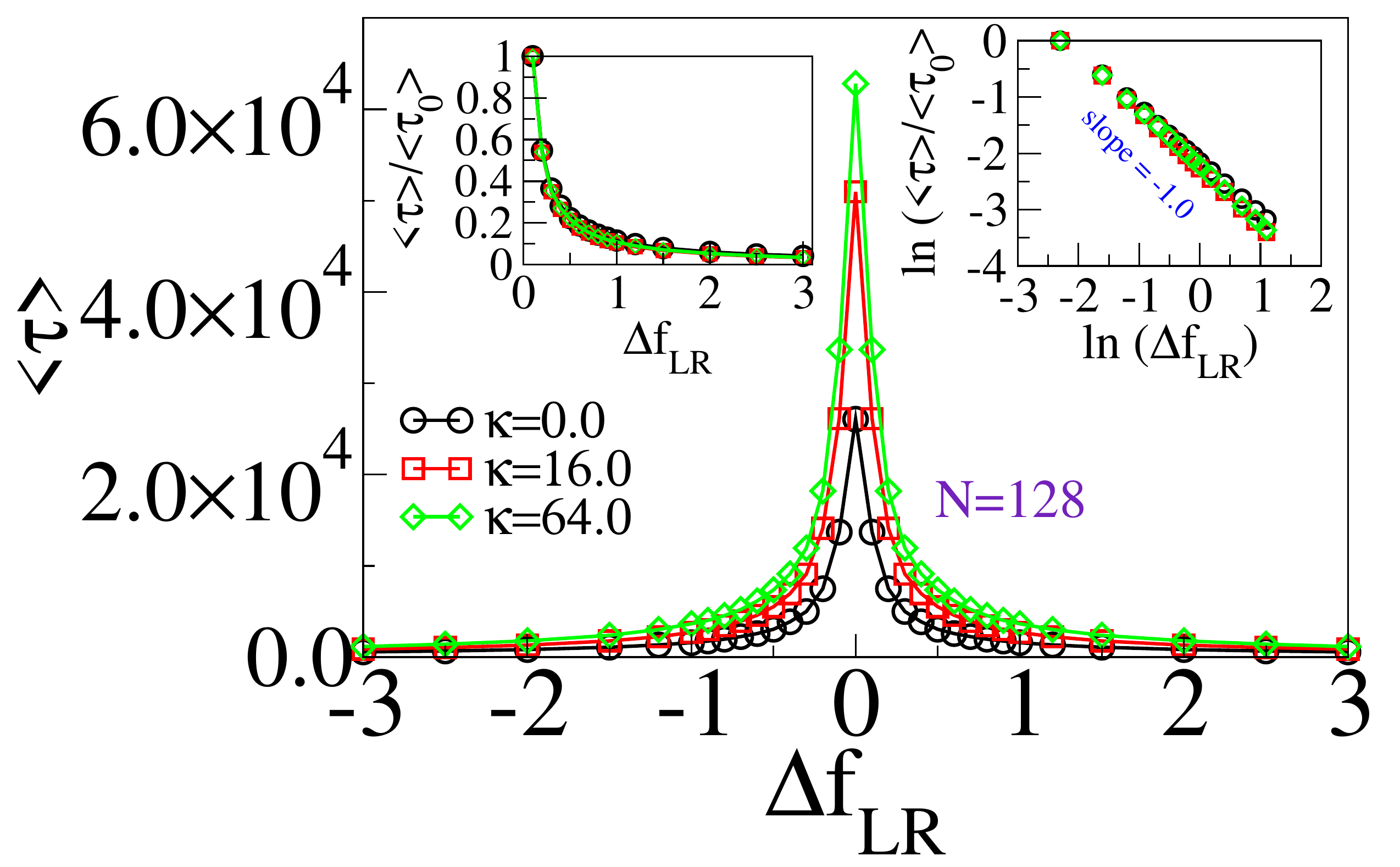}
\caption{\label{biased_MFPT}\small MFPT as a function of $\Delta  
  f_{LR}$. The black, red, green, and blue lines refer to chain  
  stiffness $\kappa = 0$, 16, and 64 respectively. The inset at  
  the left  shows normalized MFPT where all the plots collapse onto  
  the same master plot. The inset at the right is the corresponding  
  plot of the inset at left on a logarithmic scale.}
\end{figure}

In a TOW situation, the translocation process is diffusive and hence slow. Thus a more
desirable situation is to apply a net, albeit a small bias 
$\Delta{\vec{f}_{LR} }= \vec{f}_L + \vec{f}_R \ne 0$ so that the DNA
can move slowly. The presence of two forces at each pore can provide a
feedback mechanism to 
control the movement of the translocating chain~\cite{TwoPore2}.
We define $\Delta f_{LR}=\pm \left|\Delta f_{LR} \right|$, where the $\pm$
sign refers to direction of the net force $\vec{f}_{LR}$ (positive/negative
for left/right to right/left translocation). We observe that for a low bias ($\left|\Delta f_{LR} \right| <  k_BT/
\sigma$),  the MFPT
initially decreases almost exponentially (Fig.~\ref{biased_MFPT}), and
when the bias is increased beyond $\left|\Delta f_{LR}\right|\sigma  \ge  k_BT$,
then the MFPT decays with a power law $\langle \tau \rangle \sim |\Delta
f_{LR} |^{-1} $. The shape of the curve from our simulation (Fig.~\ref{biased_MFPT}) is almost the same as reported in DNP experiment  by Zhang
{\em et al.}~\cite{TwoPore2}.  Since the length and time scale of simulation and experimental
scales are different, it seems that this behavior is generic, 
independent of the size of the system.
The inset of Fig.~\ref{biased_MFPT} at the left shows the normalized $\langle \tau \rangle
/\langle \tau_0 \rangle$ 
for several values of $\kappa$, 
demonstrating that this is a generic feature for a wide range of chain
stiffness. Here we have chosen the normalization factor $\langle
\tau_0 \rangle$ to be
the MFPT for $\Delta  f_{LR}=0.1$. This eliminates the chain length dependence
of $\langle \tau \rangle$. 
The inset Fig.~\ref{biased_MFPT} at the right shows that
the $\langle \tau \rangle \sim |\Delta f_{LR} |^{-1}$ (for $1.0 <
\Delta f_{LR} \le 3$) ~\cite{Adhikari_JCP_2013}. We have checked that
for $d_{LR}=16$ and 24 and for chain length $N=64 - 256$ that this trend is the same. 
Similar dependence on the force has been observed for SNP
translocation~\cite{Adhikari_JCP_2013}.
\par
\begin{figure}[ht!]
\includegraphics[width=0.45\textwidth]{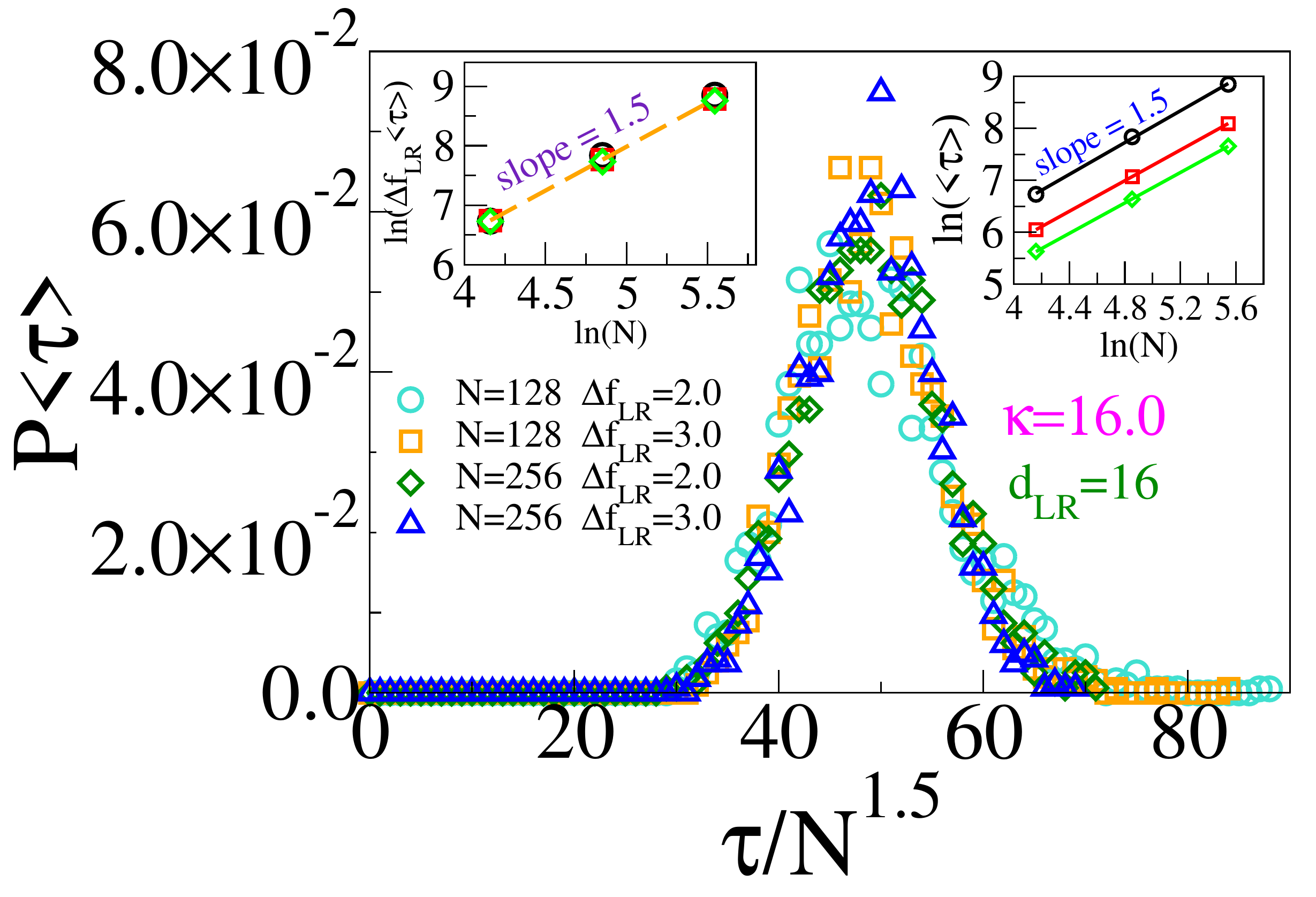}
\caption{\label{scaled_MFPT} \small 
Histogram of the 
  MFPT as a function of $\tau/N^{1.5}$ for two  different net bias $\Delta F_{LR}$  and two 
  chain length $N=128$ and 256 which shows data collapse; Teal circles 
  and orange squares correspond to $N=128$ and  $\Delta F_{LR}=2$ and 
  3.  Green diamonds and blue circles correspond to $N=256$ and  $\Delta F_{LR}=2$ and 
  3.  The inset at the right shows plots for $\langle \tau \rangle 
  \sim N $ (log scale) for $\Delta F_{LR}=1$ (black circles), 2 (red squares), and 
  3 (green diamonds). The straight lines in each case is a linear fit 
  with slope $1.5 \pm 0.02$. The inset at the left is the 
  corresponding plot of $\Delta F_{LR}   \langle \tau \rangle \sim N $.}
\end{figure}

We complete our scaling analysis by studying the chain length
dependence of the MFPT. We find  that
$\langle \tau \rangle \sim N^{1.5}$ as evident from (i) the data collapse
of the histogram of the MFPT and the (ii) two
insets of Fig.~\ref{scaled_MFPT}.  The slope for each curve at the
inset at the right for $\Delta f_{LR}= 1, 2, 3$ is $1.5 \pm 0.02$. 
Fig.~\ref{scaled_MFPT} shows this data collapse. Furthermore,
combining these results with $ \langle \tau \rangle \sim |\Delta f_{LR}
|^{-1}$ we obtain
\begin{equation}
 \label{scale}
 \langle \tau \rangle = A|\Delta f_{LR}|^{-1}N^{1.5}
\end{equation}
The inset at the left of Fig.~\ref{scaled_MFPT} convincingly shows the data collapse for two chain lengths $N=128$
and 256 for two values of  $\Delta f_{LR}$ = 2 and 3 respectively,
verifying Eqn.~\ref{scale}.
This power law scaling of Eqn.~\ref{scale} with the value of the
effective translocation exponent $\alpha \approx
1.5 $ for chain lengths $N\sim 100-500$ is the  
same as observed in a biased
SNP~\cite{Bhattacharya_Proceedia, Kaifu_PRE2008,Bhattacharya_EPJE}. This is discussed in detail below. \par
The scaling ansatz for the MFPT of a fully flexible chain in the
context of a SNP translocation is
given by~\cite{Ikonen_EPL2013,Ikonen_JCP2012}
\begin{equation}
 \langle \tau \rangle  \sim [ AN^{1+\nu} + \eta_{pore} N ] \left |\Delta
 F_{LR} \right |^{-1} \sim N^{\alpha}.
\label{ansatz}
\end{equation}
Here $A$ is constant, $\nu=0.75$ and 0.5888 is the Flory exponent in 2D and
3D, respectively, and $\eta_{pore}$ is the pore friction. 
Eqn.~\ref{ansatz} for a self-avoiding fully flexible chain
holds for small to strong stretching force limits (trumpet,
stem-flower, and strongly stretched) regimes~\cite{Ikonen_EPL2013,Ikonen_JCP2012}.\par
Following Cantor and Kardar, the origin of the first term is $\langle \tau \rangle \sim
\langle R_g \rangle /v_{CM} = N^\nu/N = N^{1+\nu}$~\cite{Cantor-Kardar}, where $\langle R_g \rangle$ is the
radius of gyration of the chain.
The second term
$\eta_{pore}N$ in Eqn.~\ref{ansatz} is the contribution of
the pore friction $\eta_{pore}$ and proportional to the contour length of
the chain.  When the chain length is small, then the pore
friction term has a significant effect on the effective translocation
exponent $\alpha$ and $1<\alpha<1+\nu$~\cite{Bhattacharya_Proceedia,Bhattacharya_EPJE}. This is the reason for the
smaller value of the effective translocation exponent $1.5$ for the range of $100 < N < 500$ rather than the
$\alpha=1+\nu = 1.75$ (2D). In the long chain limit $\alpha \rightarrow 1+\nu$, as
the dominant contribution to the translocation time comes from the
friction due to the movement of the polymer inside the solvent.\par
\begin{figure}[ht!]
\includegraphics[width=0.45\textwidth]{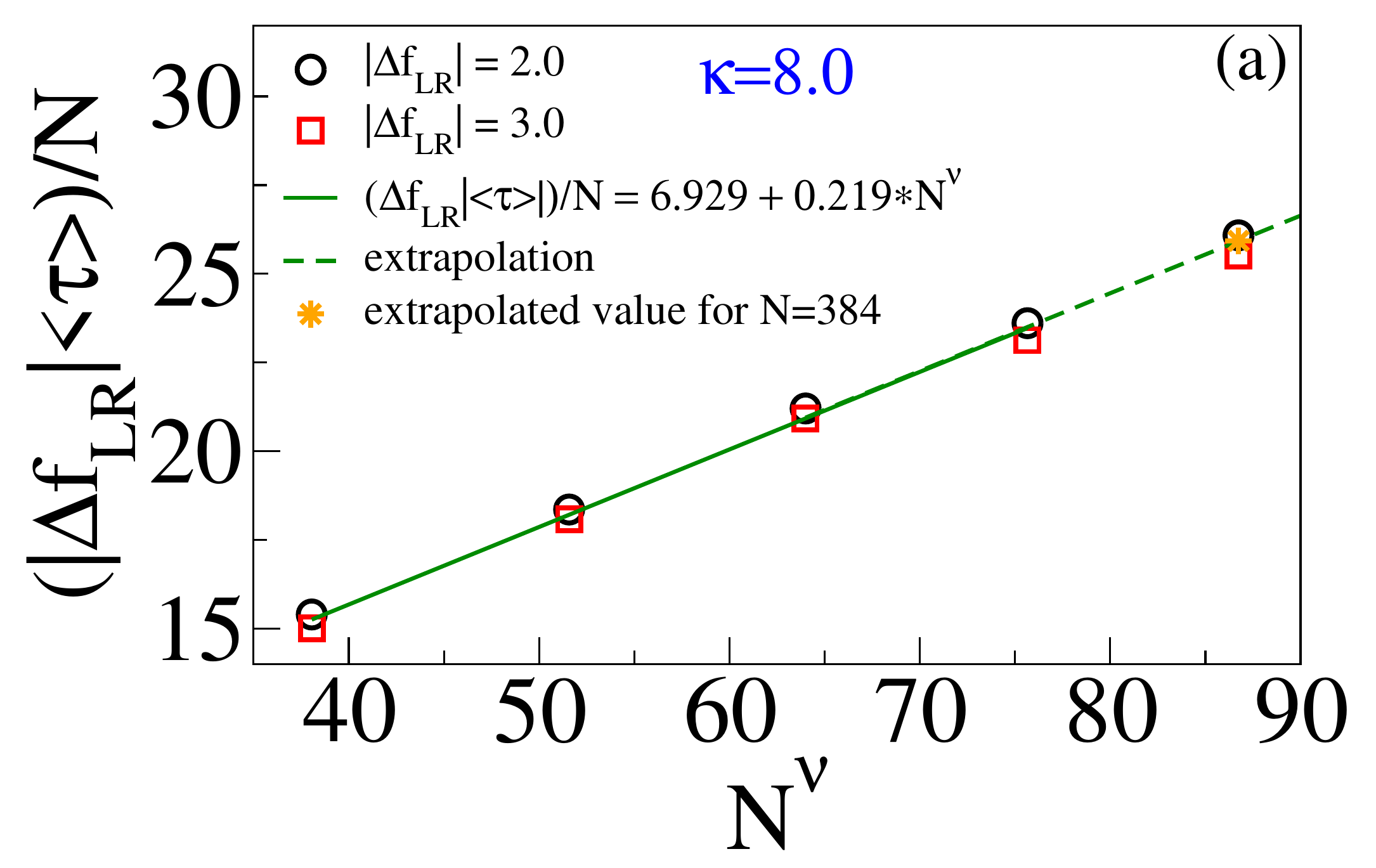}\\
\includegraphics[width=0.45\textwidth]{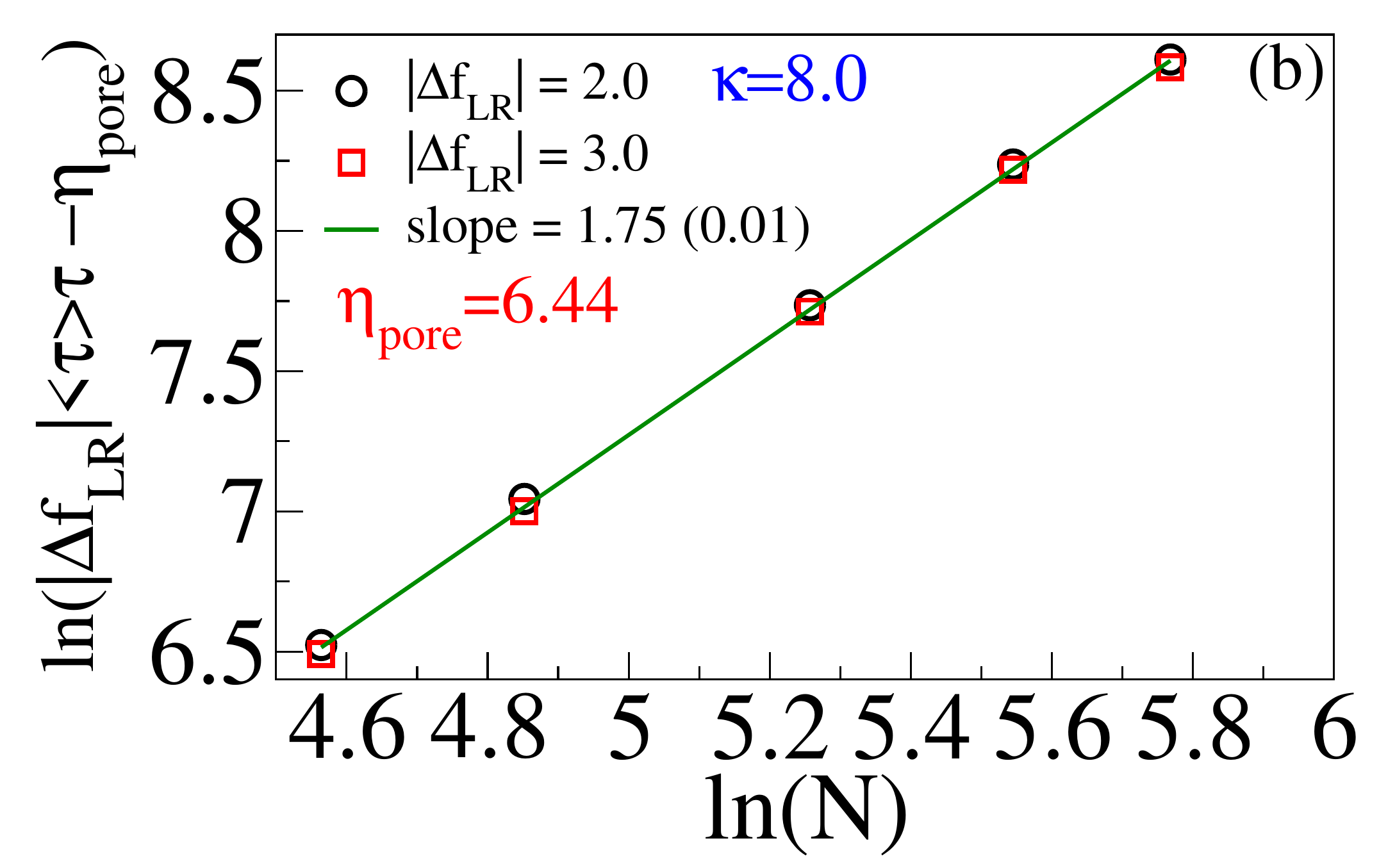}\\
\caption{\label{finite-size} \small  
(a) Plot of   $\langle \tau \rangle \left|\Delta f_{LR}\right |/N$  as a  
function of $N^\nu$ for $\left|\Delta f_{LR}\right | = 2$ (black circles)  
and 3 (red squares). The green solid line is a straight  
line fit through the points (which are average of black circles and  
red squares for each value of $N=128,192, 256$, and 320 respectively) enables us to
determine  $A=0.219$ and 
$\eta_{pore}=6.929$.  The brown star corresponds to the predicted value for
$N=384$ from the extrapolated line (green dashed line), while the black circle and red square are the data obtained
from simulation.  (b) Demonstration that subtracting the pore friction  
one regains the asymptotic exponent of $1+\nu \approx 1.75 \pm 0.01$ for long chains.}
\label{extrapolate}
\end{figure}

When the chain persistent length $\ell_p<<L$, the Flory theory for the
radius of gyration following Nakanishi~\cite{Nakanishi} and Schaeffer,
Pincus and Joanny~\cite{Pincus_MM_1980} is written as
\begin{equation}
\sqrt{\langle
  R_g^2\rangle} \sim \ell_p^{1/d+2}N^\nu.
\label{saw}
\end{equation}
In a previous paper we have shown the regime of $L/\ell_p$ where the
above relation is strictly valid~\cite{Huang_JCP_2014}.
Thus, in the limit when $\ell_p<<L$,  a plausible generalization of 
the scaling ansatz of Eqn.~\ref{ansatz} is 
\begin{equation}
 \langle \tau \rangle  \sim [ A^\prime \ell_p^{1/d+2}N^{1+\nu} +
 \tilde{\eta}_{pore}(\ell_p) N ] \left |\Delta
 F_{LR} \right|^{-1} \sim N^{\alpha}.
 \label{ansatz2}
 \end{equation}
The second term
$\tilde{\eta}_{pore}(\ell_p)$ is now is a function of the chain persistence
length. Thus for a given persistence length $\ell_p$, 
Eqn.~\ref{ansatz2} reduces to Eqn.~\ref{ansatz} with $A
=\ell_p^{1/d+2}A^\prime$, which then can be used  to predict the behavior of
longer semiflexible chains for the cases when $\ell_p << L$. The MFPT
for shorter chains (still for those chain lengths for which
$L/\ell_p>> 1$) are used to numerically obtain $A$ and $\eta_{pore}(\ell_p)$, which
then can be substituted in to Eqn.~\ref{ansatz} to predict the MFPT of longer
chains. This scheme is shown in
Fig.~\ref{extrapolate}. 
Eqn.~\ref{ansatz} can be rewritten as follows: 
\begin{equation}
\frac{\langle \tau \rangle \left|
  \Delta f_{LR}\right |}{N} = AN^\nu + \eta_{pore}(\ell_p) 
 \end{equation}
Thus for a given value of the chain 
persistence length $\ell_p$, a plot of   $\langle \tau \rangle \left|
  \Delta f_{LR}\right |/N$  as a 
function of $N^\nu$ will produce a straight line with the slope $A$ and  
$\eta_{pore}$ as the intercept. 
Fig.~\ref{extrapolate}(a) clearly shows the validity of the scaling
ansatz as in 
Eqn.~\ref{ansatz}. Here we have used Eqn.~\ref{ansatz2} and the simulation
data for the MFPT for chain lengths $N=128, 192, 256$, and 320  to obtain the
value of the pore friction $\tilde{\eta}_{pore}$ for $\kappa =
8.0$ (the corresponding $\ell_p = 2\kappa /k_BT = 13.3$). The choice
of these chain lengths satisfy $2.2 < L/\ell_p < 3.2$, 
and from our previous work~\cite{Huang_JCP_2014} we confirm 
that these combinations of $L$ and $\ell_p$ are well in the two
dimensional self-avoiding random walk (2DSAW) regime and satisfy
Eqn.~\ref{saw}. 
A linear regression perfectly fits the simulation data and that the data for different 
values of the bias $\left|\Delta f_{LR}\right | = 2$ and 3 collapse on the same 
master plot with $A = 0.219$ and $\eta_{pore} = 6.929$.
Substituting these values of $A$ and $\eta_{pore}$ into Eqn
~\ref{ansatz} (${\langle \tau \rangle \left | \Delta f_{LR}\right |}/{N} = 0.219N^\nu + 6.929$)  we 
extrapolate (dashed green line in Fig.~\ref{extrapolate}(a)) to 
predict the MFPT for $N=384$ (brown star) and check that this point
falls on top of the 
simulation data for $N=384$ (black circles and red squares).  
Once $\eta_{pore}$ is known, one can also check that 
subtracting the pore friction contribution from $\langle \tau \rangle$
provides a slope of $1+\nu$. This is shown in 
Fig.~\ref{extrapolate}(b) where a log-log plot of
$\langle \tau \rangle \left|\Delta f_{LR}\right | - \eta_{pore}N$ versus $N$ indeed produces a slope 
$\approx 1+\nu$ (the regression produces a slope $1.75\pm 0.01$). This
proves that in the limit $d_{LR}/L << 1$ the scaling ansatz for the
SNP translocation works for the model DNP system.\par
It is worth noting that generalizing the scaling ansatz in three dimensions (3D)
will be more challenging as unlike in 2D where the Gaussian regime is
absent~\cite{Huang_JCP_2014}, the self-avoiding random walk (SAW) in
3D (3DSAW) appears at the end of the Gaussian regime and hence will
require much longer chain lengths. Additionally, it is worth
investigating the scaling ansatzs for different regimes (rod, Gaussian
and SAW) and for larger biases when the additional friction from the
translocated segment of the chain may need to be incorporated separately~\cite{Jalal_2017}.
\section{Summary and Conclusion}
To conclude, we have studied various aspects of translocation in a
model DNP system. The system we have studied is an ideal system
motivated by recent experiments to answer some  general characteristics
of translocation through a DNP in the limit $d_{LR}/L << 1$. 
Our studies of the TOW shows that the
effect of the magnitude of the TOW forces will be more prominent for a ssDNA but likely to be
insignificant for long dsDNA segments used in recent
experiments. These conclusions were verified by studying the
translocation of a triblock copolymer ABA and using the known result
that a stiffer polymer translocates slower through a SNP. One of the
primary motivation of recent
experiments is to measure the current blockade
time (TOF) for a tagged DNA segment of known length 
translocating through a DNP system. If the segment moves through
the DNP with a 
constant velocity, then the current blockade time can be readily
translated to the corresponding genomic length.
Thus our studies of TOF is directly
 relevant for DNP experiments where the goal is to extract genomic
 distances from the data obtained in the time domain. We demonstrate that the TOF has a
 quasi-periodic structure, (implicating {\em non-uniform
 speed}).  In the limit when $d_{LR} << L$ by applying a
 net bias on the DNP we demonstrate that we recover scaling laws of
 SNP translocation. When $d_{LR}/ L << 1$ and the chain is subject to
 TOW forces, the entropic contribution from the segment in between the
 pore is almost insignificant compared to the total entropy of the
 chain. Thus, in this limit the pore friction term $\eta_{pore}N$
 from each pore adds up linearly (so that for the case of pores of
 same width it will simply be proportional to the number of pores), and one recovers
 the scaling ansatz for the SNP translocation. A DNP is an 
interesting system where one can 
create different chain conformations with variable tension and
stiffness for the chain segment in between the pores by adjusting 
$\vec{f}_L$, $\vec{f}_L$, and $d_{LR}$. Thus simulation studies of
block copolymers and random heteropolymer 
translocation through DNP systems can produce intriguing and exciting results for studying nonlinear elasticity 
of biopolymers~\cite{Janmay,Dobrynin1,Dobrynin2}. We hope that these
results will provide further insights to design new experiments, be
useful for making a theoretical framework for multi-pore
translocation, and 
promote further work in this direction. \\
\section{Acknowledgment}
The authors acknowledge computing resources under the auspices of
UCF's high performance computing cluster STOKES where all the
computations were done. AB thanks Profs. Kurt Binder and Walter
Reisner for various discussion and comments on the manuscripts. The
authors gratefully acknowledge and thank both the referees 
for their comments and critiques on the manuscript. 


\medskip
\begin{thebibliography}{29}
\bibitem{Bashir}B.~M. Venkatesan and R. Bashir, Nature Nanotechnology
  {\bf 6}, 615–624 (2011).

\bibitem{Review}For recent reviews in the field please see 
M. Muthukumar \textit{Polymer Translocation} (CRC Press, Boca Raton, 2011); 
A. Milchev, J. Phys. Condens. Matter {\bf 23}, 103101 (2011);
D. Panja, J. Stat. Mech. P06011 (2010);  
Vladimir V. Palyulin, Tapio Ala-Nissila,  and Ralf Metzler, Soft Matter, {\bf 10}, 9016 (2014). 
  
\bibitem{Kasianowicz} J. J. Kasianowicz, E. Brandin, D. Branton and D. W. Deamer, \textit{Proc. Natl. Acad. Sci. U.S.A.} {\bf 93}, 13770 (1996).
\bibitem{Meller00} A. Meller, L. Nivon, E. Brandin, J. A. Golovchenko, and D. Branton, \textit{Proc. Natl. Acad. Sci. U.S.A.} {\bf 97}, 1079 (2000).
\bibitem{Meller01} A. Meller, L. Nivon, and D. Branton, \textit{Phys. Rev. Lett.} {\bf 86}, 3435 (2001).

\bibitem{Meller02} A. Meller and D. Branton, \textit{Electrophoresis} {\bf 23}, 2583 (2002).

\bibitem{Dekker}S. Pud, S. Chao, M. Belkin, D. Verschureren, 
  T. Huijben, C. van Engelenburg, C. Dekker, and A. Aksimentiev, Nano 
  Lett.  {\bf 16}, 8021 (2016). 

\bibitem{TwoPore1} Y. Zhang, X. Liu, Y.  Zhao, J. -K. Yu, W.
  Reisner, and W. B. Dunbar, Small {\bf 14}, 1801890 (2018).

\bibitem{TwoPore2} X. Liu, Y. Zhang, R. Nagel, 
  W. Reisner, W.~B. Dunbar, Small {\bf 15}, 1901704 (2019).

\bibitem{Flossing}X. Liu, P. Zimny, Y. Zhang, A. Rana, R. Nagel, W. Reisner,
and W.~B. Dunbar, Small {\bf 16}, 1905379 (2020).

\bibitem{Langecker}M. Langecker, Nano Lett. {\bf 11}, 5002 (2011). 

\bibitem{Cadinu1}Paolo Cadinu {\em et al.} Nano Lett. {\bf 17}, 6376 (2017). 

 \bibitem{Cadinu2}Paolo Cadinu {\em et al.} Nano Lett. {\bf 18}, 2738 (2018).
   
\bibitem{Briggs}K. Briggs {\em et al.}, Nano Lett. {\bf 18}, 660 (2018). 

\bibitem{Yeh}Jia-Wei Yeh, Nano Lett. {\bf 12}, 1597 (2012). 

\bibitem{Ikonen_EPL2013} 
Ikonen T,  Bhattacharya A,  Ala-Nissila T,  and Sung W 
{\it EPL}{\bf 103} 38001 (2013).

\bibitem{Ikonen_JCP2012}
 Ikonen T , Bhattacharya A.,  Ala-Nissila T. and Sung W. {\it 
   J. Chem. Phys.}{\bf 137} 085101 (2012). 

\bibitem{Jalal_2017}J. Sarabadoni, T. Ikonen, H. Mokk$\ddot{\rm o}$k$\ddot{\rm o}$ken, 
  T. Ala-Nissila, S. Carson,   M. Wanunu, Sci. Rep. {\bf 7}, 7423
  (2017).
  
\bibitem{Ikonen_PRE2012} T. Ikonen, A. Bhattacharya, T. Ala-Nissila and W. Sung, Phys. Rev. E 
{\bf 85}, 051803 (2012). 

\bibitem{Sakaue_PRE_2007}T. Sakaue, {\it Phys. Rev. E} {\bf 76} 021803
  (2007); {\it ibid} {\bf 81}, 040808 (2010);   T. Saito and T. Sakaue, Eur. Phys. J. E {\bf 34}, 135
  (2011).

\bibitem{Adhikari_JCP_2013}
 Adhikari R. and Bhattacharya A. {\it J. Chem. Phys.} {\bf 138}, 204909 (2013).

\bibitem{Grest}G.~S.Grest and K. Kremer, Phys. Rev. A \textbf{1986}, \textit{33}, 3628(R) (1986).

\bibitem{Binder_Review}A. D. Sokal, in \textit{Monte Carlo and Molecular Dynamics Simulations in Polymer 
Science}, edited by K. Binder (Oxford University Press, New York, 1995), Chap. 2 

\bibitem{Huang_EPL_2014a}
A. Huang, R. Adhikari, A. Bhattacharya, and K. Binder, 
\textit{Europhys. Lett.} \textbf{105}, 18002 (2014). 

\bibitem{Huang_EPL_2014b}
A. Huang and A. Bhattacharya, \emph{Europhys. Lett.}
\textbf{106}, 18004 (2014).

\bibitem{Huang_JCP_2014}
A. Huang, A. Bhattacharya, and K. Binder, \emph{J. Chem. Phys.} \textbf{140}, 214902 (2014). 

\bibitem{Huang_JCP_2015} A. Huang, H.-P. Hsu, A.  Bhattacharya, and
  K. Binder, \textit{J. Chem. Phys.} \textbf{143}, 243102 (2015). 

\bibitem{Polymers2016} A. Huang, W. Reisner, and A. Bhattacharya,
Polymers \textbf{8}, 352 (2016).

\bibitem{MM2018}S. Bernier S, A. Huang, W. Reisner, and A. Bhattacharya, Macromolecules {\bf 51}, 4012 (2018). 

\bibitem{Kaifu_PRL}K. Luo, T. Ala-Nissila, S. -C. Ying, and A. Bhattacharya,
Phys. Rev. Lett. {\bf 99}, 148102 (2007); \textit{ibid.} {\bf 100},
058101 (2008).

\bibitem{Bhattacharya_EPJE} A. Bhattacharya, W.H. Morrison, K. Luo,
  T. Ala-Nissila, S. -C. Ying, 
A. Milchev and K. Binder, Eur. Phys. J. E {\bf 29}, 423-429 (2009).

\bibitem{Bhattacharya2010} A. Bhattacharya and K. Binder, Phys. Rev. E {\bf 81}, 041804 (2010).

  \bibitem{Bhattacharya_Proceedia}A. Bhattacharya,  
Computer Simulation Studies in Condensed Matter Physics XXII, 
Eds. D. P. Landau, S. P. Lewis, and H. B. Schuttler, Elsevier, Physics 
Proceedia {\bf 3}, 1411 (2010). 

\bibitem{Landau} L.~D. Landau and E.~M. Lifshitz, \textit{Statistical
    Physics}, Part 1, 3rd ed. (Pergamon Press, 1980).
 
\bibitem{Langevein} van Gunsteren, W. F.; Berendsen, H. J. C.   \textit{Mol. Phys.} \textbf{1982}, \textit{45}, 637. 

  \bibitem{Allen} M.~P. Allen and D.~J. Tildessley, \textit{Computer
      Simulation of Liquids} (Oxford University Press, Oxford,1987). 
    
 \bibitem{Rubinstein}M. Rubinstein and Ralph H. Colby, \textit{Polymer Physics}, (Oxford University Press, 2003).   

   
\bibitem{Cantor-Kardar}Y. Kantor and M. Kardar, Phys. Rev. E {\bf 69},
  021806 (2004).

\bibitem{Nakanishi}H. Nakanishi, J. Phys. {\bf 48}, 979 (1987); 
  J. Moon and H. Nakanishi, Phys. Rev. A {\bf 44}, 6427 (1991).
  
\bibitem{Pincus_MM_1980}D.~W. Schaefer, J. F. Joanny, and P. Pincus,
  Macromolecules {\bf 13}, 1280 (1980).

\bibitem{Kaifu_PRE2008}K. Luo, S. Ollila, I. Huopaniemi, T. Ala-Nissila, P. Pomorski, M. Karttunen, 
S-C. Ying, and A. Bhattacharya, Phys. Rev. E. {\bf 78}, 050901(R)
(2008).

\bibitem{Janmay}Cornelis Storm, Jennifer J. Pastore, F. C. MacKintosh,
T. C. Lubensky, \& Paul A. Janmay, Nature {\bf 435}, 191 (2005).

\bibitem{Dobrynin1}Andrey V. Dobrynin and Jan-Michael Y. Carrillo, Macromolecules {\bf 44}, 140 (2011).

\bibitem{Dobrynin2}Jan-Michael Y. Carrillo, Fred C. MacKintosh, and
  Andrey V. Dobrynin, Macromolecules {\bf 46}, 3679 (2013).

\end{thebibliography}
\end{document}